\documentclass{appolb}
\usepackage{epsfig}
\newcommand{\mathbb}[1]{I\!\!{#1}}
\begin{document}
\title{Dark energy and 3-manifold topology%
\thanks{Presented at the conference ``Matter to the deepest'' Ustron 2007}%
}
\author{Torsten Asselmeyer-Maluga\thanks{torsten.asselmeyer-maluga@dlr.de}
\address{German Aerospace Center, Rutherfordstr. 2, 12489 Berlin, Germany}
\and
Helge Ros\'{e}\thanks{rose@first.fhg.de}
\address{Fraunhofer FIRST, Kekulestr. 7, 12489 Berlin, Germany}
}
\maketitle
\begin{abstract}
We show that the differential-geometric description of matter by
differential structures of spacetime leads to a unifying model of the
three types of energy in the cosmos: matter, dark matter and dark
energy. Using this model we are able to calculate the ratio of dark
energy to the total energy of the cosmos.
\end{abstract}
  
\section{Introduction}

For centuries it has been our firm conviction that matter and energy
of the same kind as is surrounding us also constitute the rest of the
world. Thorough examinations of supernovae
\cite{nova:99,nova:05} and of cosmic background radiation
\cite{WMAP:03:1,WMAP:03:2}, however, have replaced this
conviction by the insight that the global structure of the cosmos is
dominated at 95\,\% by an energy form that has hitherto been entirely
unknown. About two thirds of this energy form consist in {}``dark
energy'', and one third in {}``dark matter''. This is the most radical
revolution in our understanding of the cosmos after Kopernikus. In the
last years, great effort has been invested to understand these unknown
forms of energy \cite{cosmo:03:1,cosmo:03,cosmo:04}.  Many
explanations of dark energy assume that besides spacetime geometry and
baryonic matter, there is an \emph{additional entity} that acts as
source of the dark energy. For instance, particle-theoretic models
attribute this role to the vacuum energy
\cite{lambda:67,lambda:89,lambda:01}, or introduce additional global
scalar fields \cite{quint:88,quint:88:2}.

A first evaluation of the WMAP data favors a Poincar\'{e} sphere as
topology of the cosmos \cite{dodecaeder:03}. That means that the
cosmos is a closed 3-manifold having the same homology as the
3-sphere. Furthermore one knows that this 3-manifold has a positive
curvature. In this paper we support this result and show that the dark
energy can be explained by the curvature of two Poincar\'{e}
spheres. Details of the calculation can be found in the expanded version of the paper \cite{AssRos:06}.

\section{Basic Model}
Our model based on the fact from general relativity theory that every
form of energy is related to the curvature of the spatial 3-manifold,
i.e. matter must be interpreted by curved 3-manifolds. Then Einstein's
equation is the dynamical equation for the evolution of 3-geometries.
Thus we can state our basic assumptions:\\ \textbf{Basic assumptions}:
\emph{The 4-manifold of all possible spacetime events is a compact,
  closed 4-manifold} $M$ \emph{which is differentiable and simply
  connected.  The cosmos is an embedded 3-manifold } $\Sigma$
\emph{which is compact and closed.  The energy density of any kind of
  matter is described by the curvature of the associated submanifold
  of}~$\Sigma$.

Before we study the implications we will motivate these
assumptions. Compactness\footnote{Compactness is not contradictory to
  the possible infiniteness of proper time for world lines. Curves in
  manifolds can have infinite length like Peano´s curve.} of the
4-manifolds means that \emph{every} serie of spacetime events
converges to an event belonging to the same 4-manifold. The manifold
is closed i.e. it has no boundary or in any neighborhood of an
arbitrary point there is always inner points of the manifold. The
assumptions of compactness and closedness can be interpreted that all
points of the manifold are inner points and any spacetime event must
be part of the manifold. That assumption seems natural from our
knowledge of space and time. The assumption of simple-connectness is
more delicate. It means that any closed (time-like) curve is
contractable i.e. any time circle contradicting causality can be
shrunk to a point.  In the following we will study the implications
via differential topology of the assumptions leading to very strong
restrictions on the possible 4- and 3-manifolds.
\subsection{Determination of the 4-manifold}
The 4-manifold can be determined by the following argumentation. If
every kind of energy is described by the curvature of some submanifold
of the spatial 3-manifold $\Sigma$ then there are no source terms in
Einstein's equation. Every kind of energy including matter must be
given by geometry. Then Einstein's equation
$R_{\mu\nu}=0$
is the statement that the 4-manifold must be Ricci-flat. But
\emph{there is only one Ricci-flat compact, simply-connected
 4-manifold, the K3-surface} \cite{Yau:77} which is defined by
\[
\{(z_1,z_2,z_3,z_4)\in\mathbb{\!\!C}P^3\, |\,
z_1^4+z_2^4+z_3^4+z_4^4=0\}\quad .
\]
Thus we determine the topology of the 4-manifold by our basic
assumptions. The second restriction, which will determine the structure
of the 3-manifold, is the choice of a particular differential
structure. It is a well-known result \cite{GomSti:1999} that the
differential or smooth structure of a compact, simply-connected
4-manifold is determined by a contractable submanifold, the Akbulut
cork~$A$. The boundary of the Akbulut cork is a 3-manifold which is
a so-called \emph{homology 3-sphere}, i.e. a 3-manifold with the same
homology as the 3-sphere $S^3$. Thus we assume
\emph{the cosmos $\Sigma$ is a homology 3-sphere}. 
\subsection{The 3-manifold and its submanifolds}
In case of the K3-surface we know the structure of the Akbulut cork
and its boundary $\Sigma=\Sigma(2,5,7)$ \cite{GomSti:1999} the Brieskorn
sphere
\[\Sigma(2,5,7)=\{(x,y,z)\in\mathbb{\!\!{C}}^3\,|\,
|x|^2+|y|^2+|z|^2=1,x^2+y^5+z^7=0\}\, . \]
From the structure theory
of 3-manifolds \cite{JacSha:79,Thu:97} we know that there are only
\emph{three} kinds of 3-manifolds that can form~$\Sigma$. In the
particular case we obtain 
\begin{equation}
\Sigma=K_{1}\#K_2\# K_{3}\#_{N}S^{3}\# S^{3}/I^{*}\# S^{3}/I^{*}.
\label{Struktur-Brieskorn}
\end{equation}
with $\#$ as the connected sum between manifolds. The symbol $S^3/I^*$
represents the Poincar{\'e} sphere which forms the global structure of
the cosmos \cite{dodecaeder:03}. Now we identify the pieces with
\begin{enumerate}
\item negatively curved pieces $K_{i}$ (matter, radiation) 
\item positively curved 3-spheres $S^{3}$ (dark matter) 
\item two positively curved Poincar\'{e} spheres $S^3/I^*$ (dark energy), 
\end{enumerate}
The details can be found in the expanded paper \cite{AssRos:06}. This
remarkable fact motivates the following

\textbf{Conjecture:} \emph{The three types of 3-manifolds that constitute
the cosmos as a homology 3-sphere, correspond to the three kinds of
matter: baryonic matter, dark matter, and dark energy.}

Thus we obtain a unified approach for all observed kinds of energy
densities. The global structure of the cosmos can thus be derived from
the differential geometry of spacetime itself, without additional
entities, and it is possible to compare the observed energy densities
with the curvatures of the three types of~$\Sigma$. In the next
section we will calculate the ratio of the energy densities of the
dark energy and the total energy by using a result of Witten
\cite{Wit:89.2}.

\section{Calculation of the dark energy density}

The investigation of the global characteristics of the differential
structure of spacetime led us to the result that the cosmos is a Brieskorn
sphere $\Sigma=\Sigma(2,5,7)$ split up into three types of pieces
\[
\Sigma=K_{1}\#K_2\# K_{3}\#_{N}S^{3}\# S^{3}/I^{*}\# S^{3}/I^{*}.
\]
Let us now suppose a 4-manifold $M$ with an Akbulut cork bounded by
$\Sigma$. The metric $g_{\mu\nu}$ of $M$ is given by the Einstein
equation\begin{equation}
R_{\mu\nu}-\frac{1}{2}g_{\mu\nu}\: R=\Lambda g_{\mu\nu}+\frac{8\pi G}{c^{4}}\cdot T_{\mu\nu}\label{Einstein-cosmological-constant}\end{equation}
 with the cosmological constant $\Lambda$ and the energy-momentum
tensor $T_{\mu\nu}$. The cosmological constant\begin{equation}
\Lambda=\frac{8\pi G}{c^{4}}\rho_{D}\end{equation}
 corresponds to the energy density $\rho_{D}=E_{D}/vol(\Sigma)$ of
the dark energy. In the previous section we showed that one can identify
the dark energy with the curvature of two Poincar\'{e} spheres $\Sigma_{D}=S^{3}/I^{*}\# S^{3}/I^{*}$.
Let $R$ be the scalar curvature of the cosmos $\Sigma$. Inserting
the Robertson-Walker-metric\begin{equation}
ds^{2}=c^{2}dt^{2}-a(t)^{2}h_{ik}dx^{i}dx^{j}\end{equation}
 with $\Sigma\times[0,1]$ and the scaling function $a(t)$ in (\ref{Einstein-cosmological-constant}),
one gets the Friedman equation\[
\left(\frac{\dot{a}(t)}{c\cdot a(t)}\right)^{2}+\frac{k}{a(t)^{2}}=\frac{8\pi G}{c^{4}}\frac{\rho}{3}+\frac{\Lambda}{3}=\frac{8\pi G}{c^{4}}\:\frac{(\rho+\rho_{D})}{3}\]
 with curvature $k=0,\pm1$. Then the scalar curvature of the cosmos
$\Sigma$ is $R=3k/a(t)^{2}$ and the Hubble constant is given by
$\dot{a}(t)/a(t)=H_{0}$. Thus, the relation of the total density
$\rho$ and scalar curvature $R$ of the cosmos is given by\begin{equation}
\rho=\frac{c^{4}}{8\pi G}R+\frac{3H_{0}^{2}c^{2}}{8\pi G}\,.\end{equation}
Using homogeneity and isotropy of the matter distribution we get by integration
\begin{equation}
\rho=\frac{\frac{c^{4}}{8\pi G}\int\limits _{\Sigma}R\,\sqrt{h}d^{3}x}{vol(\Sigma)}+\rho_{C}\,,\,\,\rho_{C}=\frac{3H_{0}^{2}c^{2}}{8\pi G}\label{def-dichte-curvature}\end{equation}
 with the critical density $\rho_{C}$. Replacing $\Sigma$ by $\Sigma_{D}$
we obtain in an analog way the dark energy density\begin{equation}
\rho_{D}=\frac{\frac{c^{4}}{8\pi G}\int\limits _{\Sigma_{D}}(R_{D}+R_{C})\sqrt{h}d^{3}x}{vol(\Sigma_{D})}\,,\,\, R_{C}=\frac{3H_{0}^{2}}{c^{2}}\:.\label{def-dichte-dunkeEnergie}\end{equation}

The main step of the calculation is to solve the integral, i.e.\ the
\emph{Einstein-Hilbert action of the dark energy} defined on
\emph{$\Sigma_{D}$}\[ S_{EH}(\Sigma_{D})=\int\limits
_{\Sigma_{D}}(R_{D}+R_{C})\sqrt{h}d^{3}x\:.\] Witten has discussed the
3-dimensional Einstein-Hilbert action in more detail \cite{Wit:89.2}.
He was able to derive the important result that $S_{EH}$ is related to
a pure topological property -- the \emph{Chern-Simons invariant} of
the manifold. Then one gets the simple relation between the
Chern-Simons invariant of a $SU(2)$ connection $A$ and the
Einstein-Hilbert action (see \cite{AssRos:06} for details)
\[S_{EH}(\Sigma_{D})=\frac{16\pi^{2}}{(1-R_{C}/3)^{2}}\cdot
CS(A,\Sigma_{D})\:.\] 
Using this result we are able to calculate the ratio of the energy
density of dark matter (\ref{def-dichte-dunkeEnergie}) and the total
density (\ref{def-dichte-curvature}) of the cosmos $\Sigma$ yielding\[
\frac{\rho_{D}}{\rho}=\frac{CS(A,\Sigma_{D})}{CS(A,\Sigma)}\cdot\frac{vol(\Sigma)}{vol(\Sigma_{D})}\,.\]
With the dark energy part $\Sigma_{D}=S^{3}/I^{*}\# S^{3}/I^{*}$ by
using $CS(A,M_{1}\# M_{2})=CS(A,M_{1})+CS(A,M_{2})$ we
get \begin{equation}
  \frac{\rho_{D}}{\rho}=\frac{vol(\Sigma)}{vol(\Sigma_{D})}\cdot\frac{2CS(A,S^{3}/I^{*})}{CS(A,\Sigma)}\,.\label{ratio-CS-invariants}\end{equation}
By homogeneity, the density $\rho|_{\Sigma_{D}}=E_{D}/vol(\Sigma_{D})$
restricted to the subset $\Sigma_{D}$ has to be equal to the energy
density $\rho=E/vol(\Sigma)$ on the whole manifold $\Sigma$,
i.e.\ $E_{D}/vol(\Sigma_{D})=E/vol(\Sigma)$.  With
$E/E_{D}=\rho/\rho_{D}$ it yields to
$\rho/\rho_{D}=vol(\Sigma)/vol(\Sigma_{D})$.  Inserting in
(\ref{ratio-CS-invariants}) we obtain the \emph{dark energy
  fraction}\begin{equation}
  \frac{\rho_{D}}{\rho}=\sqrt{\frac{2CS(A,S^{3}/I^{*})}{CS(A,\Sigma)}}\end{equation}
which is a \emph{purely topological invariant}.

To give an explicit expression we need the Chern-Simons invariants of
the Poincar\'{e} sphere $S^{3}/I^{*}$ and the Brieskorn sphere
$\Sigma=\Sigma(2,5,7)$. We use the general method of Fintushel and
Stern \cite{FinSte:90,KirKla:90,FreGom:91} to calculate both
invariants.  For an unique determination of the density ratio we need
a further constraint: the chosen connection must allow for a Riemann
metric, i.e.\ it has to be a Levi-Civita connection. What we need is
the so-called \emph{minimal Chern-Simons invariant} $\tau(\Sigma)$ of
a homology 3-sphere
\begin{eqnarray}
\tau(S^{3}/I^{*}) & = &
\frac{1}{120}\,,\label{tau1}\\ \tau(\Sigma(2,5,7)) & = &
\frac{9}{280}\,.\label{tau2}\end{eqnarray} This invariant corresponds
to the self-dual or anti-self-dual solutions of a $SU(2)$ gauge theory
on $\Sigma\times\mathbb{R}$. This \emph{minimization principle}
permits an unique determination of the Chern-Simons invariants and we
obtain for the dark energy fraction\begin{equation}
\frac{\rho_{D}}{\rho}=\sqrt{\frac{2\tau(S^{3}/I^{*})}{\tau(\Sigma(2,5,7))}}\label{ratio-dichten-final}\end{equation}
and thus
\begin{equation}
\frac{\rho_{D}}{\rho}=\sqrt{\frac{(2\cdot\frac{1}{120})}{(\frac{9}{280})}}=\sqrt{\frac{14}{27}}\approx0.720\,.\end{equation}
Inserting the observed total energy density
$\rho_{obs}=(1.02\pm0.02)\cdot\rho_{C}$ from the WMAP data
\cite{WMAP:03} we obtain for the \emph{dark energy
  density}\begin{equation}
  \Omega_{D}=\frac{\rho_{D}}{\rho_{C}}=0.734\pm0.014,\label{wert-kosmologische-konstante}\end{equation}
with the critical density $\rho_{C}=3H_{0}^{2}c^{2}/8\pi G$ and the
Hubble-constant $H_{0}$. Our calculated value
(\ref{wert-kosmologische-konstante}) fits very well with the currently
observed data. In particular, the fit CMB+2dFGRS+BBN
(\cite{redshift:02} Table 1) yields $(\Omega_{D})_{obs}=0.73$ and
$\Omega_{obs}=1.013$, and we obtain
$\Omega_{D}=\sqrt{14/27}\,\Omega_{obs}=0.729$. With the observed value
of the Hubble constant $(H_{0})_{obs}=(72\pm5)\,\frac{km}{s}/Mpc$ we
obtain for the cosmological constant
\[\Lambda=\sqrt{14/27}\,\frac{3(H_0)_{obs}\,\Omega_{obs}}{c^4}\approx(1.4\pm0.2)\cdot10^{-52}m^{-2}\]

We would like to emphasize that our approach deeply requires a positive
curvature of the cosmos, i.e.\ $\Omega>1$, because our proposed
topology of the cosmos -- the Brieskorn sphere -- is a closed 3-manifold
with positive curvature. This provides a strong possibility for falsification
and should be determinable by future observations.

\section{Discussion}
In the paper we show how to calculate the ratio of the dark energy
density to the total energy density of the cosmos. Although we don't
use a quantum-gravitational argumentation for the calculation, the
appearance of the Chern-Simons invariant indicates a possible relation
to quantum gravity. The exponential of the Chern-Simons invariant
(Kodama state) is a wave function in Loop quantum gravity which can be
used for cosmology. In a follow-up paper we shall deal with this
important battery of questions in more detail.


\end{document}